\documentclass[conference]{IEEEtran}
\IEEEoverridecommandlockouts
\usepackage{cite}
\usepackage{amsmath,amssymb,amsfonts}
\usepackage{algorithmic}
\usepackage{graphicx}
\usepackage{textcomp}
\usepackage{float}
\usepackage{xcolor}
\usepackage{soul}
\usepackage{multirow}
\usepackage{subfigure}

\usepackage{fancyhdr}
\usepackage{lastpage}
\begin{document}
\renewcommand{\headrulewidth}{0pt}

\title{Modeling Communication Networks in a\\Real-Time Simulation Environment for Evaluating Controls of Shipboard Power Systems}

\vspace{-3mm}

\author{

\IEEEauthorblockN {Colin Ogilvie\IEEEauthorrefmark{1}\IEEEauthorrefmark{2}, Juan Ospina\IEEEauthorrefmark{2}, Charalambos Konstantinou\IEEEauthorrefmark{1}\IEEEauthorrefmark{2}, \\ Tuyen Vu\IEEEauthorrefmark{3}, Mark Stanovich\IEEEauthorrefmark{2}, Karl Schoder\IEEEauthorrefmark{2}, Mischa Steurer\IEEEauthorrefmark{2}}

\IEEEauthorblockA{\IEEEauthorrefmark{1}Department of Electrical and Computer Engineering, FAMU-FSU College of Engineering}

\IEEEauthorblockA{\IEEEauthorrefmark{2}Center for Advanced Power Systems, Florida State University}

\IEEEauthorblockA{\IEEEauthorrefmark{3}Electrical and Computer Engineering Department, Clarkson University}

}

\maketitle
\thispagestyle{fancy}

\begin{abstract}
Interest by the U.S. Navy in the development and deployment of advanced controls in future shipboard platforms has motivated the development of the Controls Evaluation Framework (CEF) for use in investigating dynamics present in complex automated systems. This paper reports on the implementation and investigation of a communication network component within the CEF. This implementation is designed to augment the CEF's available feature set, permitting the exploration of various communication conditions on advanced control performance. Results obtained from controller hardware-in-the-loop testing are presented and analyzed to demonstrate performance characteristics pertaining to the implemented module.
\end{abstract}


\vspace{-2mm}
\section{Introduction}

Advancements in automation and control have motivated an interest by the U.S. Navy towards their inclusion in future ship systems. Automated Power and Energy Management Systems (APEMS) are of particular interest, as they can promote levels of automation and resilience that are lacking on currently deployed platforms. Integrating APEMS requires reliable communication between control components to achieve desirable performance, as adverse communication conditions can detrimentally impact connected systems~\cite{schoder2017evaluation, konstantinou2017gps}. Developing an understanding of communication conditions that can lead to undesirable system behavior is benefited by simulation.

The power systems present in U.S. Navy ships are representative of a microgrid~\cite{nguyen2019decentralized}. Accurately representing those systems requires consideration of electrical, communication, and control components. While some software tools are capable of representing all of these components within a single environment, they typically excel within a single domain. For this reason, having the ability to utilize multiple software tools is desirable. Co-simulation becomes an attractive option, as it provides both a means to couple multiple software tools with varying time-steps and a way to distribute the computational load across multiple computing platforms. The Controls Evaluation Framework (CEF) stands as an example of a co-simulation environment capable of representing various dynamics present within a U.S. Navy ship microgrid. A component the CEF is lacking relates to the capability to controllably impart undesirable communication conditions on a device under test (DUT). Addressing this limitation has motivated the development of a communication network module.

\vspace{-1mm}
\subsection{Literature Review}

Research testbeds of interest coordinate various software tools, including both real- and non-real-time simulation in addition to Controller Hardware-in-the-Loop (CHIL), with the aim of producing a re-configurable structure that can be utilized to analyze the performance of a simulated electric grid.  
Testbeds of this type are important, because they can be employed to help identify limitations in control algorithms and the hardware in which they are deployed. These research testbeds serve as points of reference to help guide the development of similar capabilities within the CEF.

In \cite{khan2020cyberphysical}, researchers developed a testbed that employs real-time power simulation, power system protection and automation devices in tandem with a communication architecture modeled within a hardware-based Software-defined Networking (SDN) device. The case study included a modified 13-bus test distribution system utilizing three SEL-751 relay devices for switch control under different detection scenarios. The SDN is used to generate forced-trip commands to one of the SEL-751 relays, with an algorithm running on a SEL-2241 Real-Time Automation Controller (RTAC) designed to detect and prevent the transmission of erroneous commands to the pertinent relay. The results of this scenario are used to demonstrate correct operation of the employed algorithm. 

In \cite{oyewumi2019isaac}, researchers developed a real-time simulation testbed that is used as another example of a co-simulation environment incorporating communication network simulation. The investigated case study utilizes a Real-Time Digital Simulator (RTDS) to simulate a 21-bus microgrid including breakers controlled by algorithms running on two connected RTACs. Control signals are communicated over Ethernet to RTACs through a RTDS Gigabit Tranceiver Network Interface (GTNET) card. An Address Resolution Protocol (ARP) poisoning attack is initiated from a general purpose computer considered as a potential intrusion vector for a compromised SCADA substation control point. The goal of this scenario was to show that attacks could be carried out, compromising the ability of controllers to relay control signals.

The final case study explored incorporates a co-simulation consisting of power and communication systems to examine the impact of communication congestion on a control algorithm \cite{cao2019realtime}. A simple 3-bus microgrid is simulated on a Real-Time Simulator (RTS) target, including a Distributed Generator (DG) at each bus terminal. Each DG control algorithm is modeled and simulated within the same RTS target as the power simulation. A communication model is developed and simulated on a workstation running OPNET. Traffic congestion is instantiated at each router, modeled within OPNET, and is set to 90\% of the available communication channel bandwidth. Results show performance degradation in the form of undesirable voltage deviations from the target voltage in addition to an increase in time it takes for the power sharing algorithm to reach consensus.

\subsection{Contributions}
A significant limitation identified in surveyed research material relates to the number of physical hardware control devices supported within simulated environments. Most research examples supporting CHIL include fewer than ten hardware controllers. The desire to accurately represent the hypothetical control environment of future navy ships motivates an interest for the support of far greater numbers of CHIL devices. Another area of limited contribution pertains to complexity of modeled communication networks. The preliminary work presented in this paper aims to contribute to existing research in the area of power and communication network modeling in the following ways: (a) power/communication simulations incorporating large numbers of controllers ($\sim$100), and (b) modeling communication network topologies that could be used to evaluate controls deployed in U.S. Navy ships.

\begin{figure}[t]
\centerline{\includegraphics[width=0.5\textwidth]{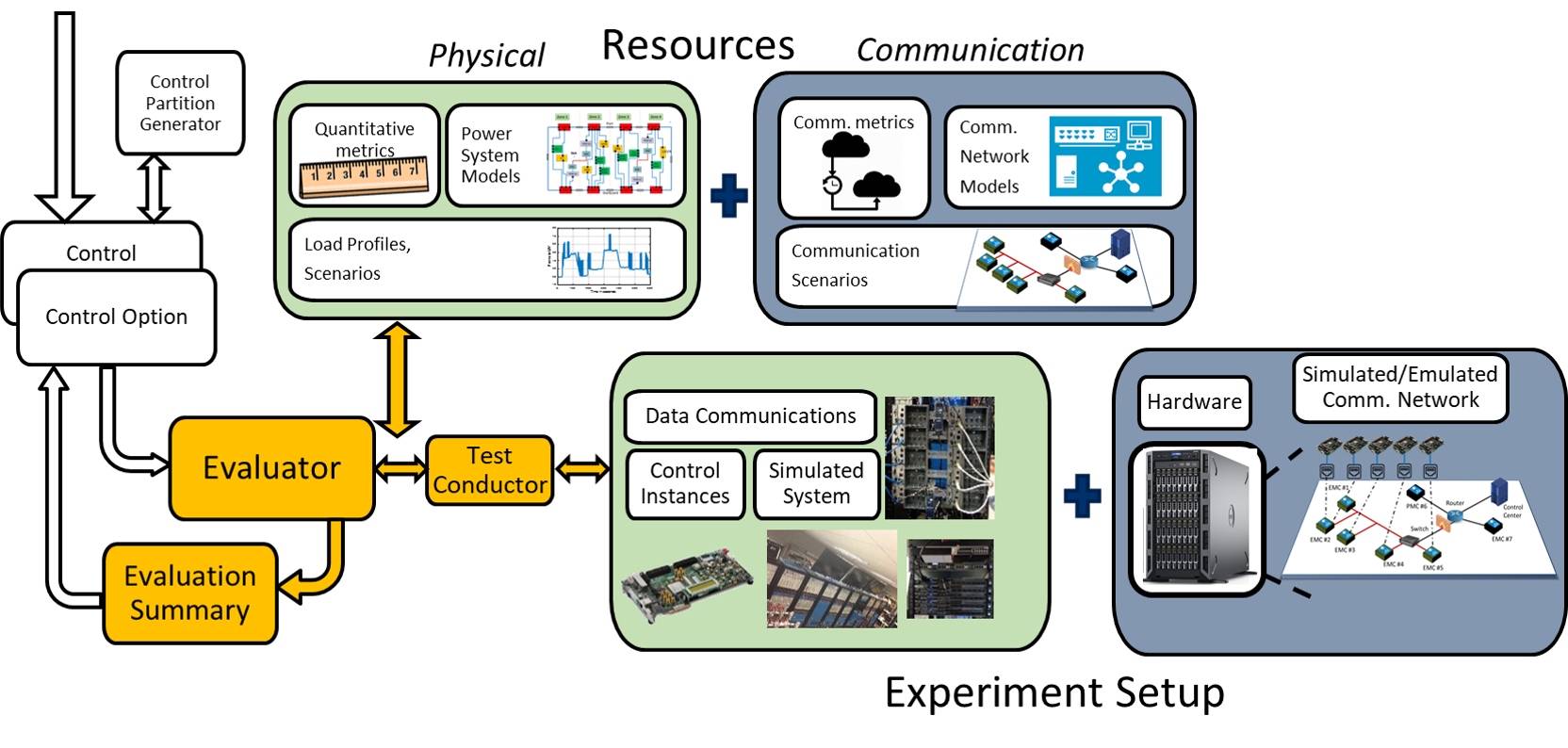}}
\caption{Block diagram of Control Evaluation Framework (CEF) with communication network component addition (blue).}
\label{fig:cef_cyber}
\end{figure}

\vspace{-1mm}
\section{Background}
\subsection{The Controls Evaluation Framework (CEF)}

As part of the Electric Ship Research and Development Consortium (ESRDC) a framework has been developed for the evaluation of system-level shipboard power systems \cite{schoder2017evaluation}. The CEF coordinates simulation hardware, controller hardware, and automated evaluation of test results to streamline control development (Fig. \ref{fig:cef_cyber}). Development of the CEF was motivated by an interest to address limitations of existing APEMS by imposing an architecture capable of simultaneously considering power system components, control system elements and their associated communication systems. APEMS encompass the various system interactions pertaining to power systems, their communication infrastructure and human interface components. Architectural implementations of APEMS take many forms, necessitating a re-configurable platform capable of exploring various configurations.

Development of APEMS requires a framework capable of supporting the deployment of a large number of networked controllers. The CEF addresses this need by incorporating a surrogate CHIL approach. This surrogate approach provides a platform for testing of distributed control algorithms at early stages of development; permitting a systematic exploration of control response and performance under controlled conditions. Several hardware and software architectures act as surrogates for control algorithm deployment, allowing comprehensive coverage of available development infrastructures. These surrogate architectures possess similar characteristics to the actual devices that will host final versions of developed control algorithms. Most importantly, the surrogate architectures employed support interfacing with real-time simulation platforms. The capability of interfacing with a real-time simulation platform is crucial, as it leads to a test setup capable of realizing the same real-time constraints imposed on end-use devices.

The communication layer of the CEF is currently representative of an ideal communication environment. While communication between simulations and surrogate platforms can take place using the same protocols and interfaces imposed on end-use devices -- the environment is not yet capable of instantiating non-ideal communication conditions, including scenarios involving statistical network effects of communication networks. It is not sufficient to assume the communication system used by end-use devices will be ideal, which necessitates a perspective that permits the analysis of non-ideal communication system behaviors. This limitation is addressed through the development of a new network emulation component that has now been integrated into the CEF.

\vspace{-1mm}
\section{Methodology \& Model Development}

\begin{figure}[t]
\centerline{\includegraphics[width=0.48\textwidth]{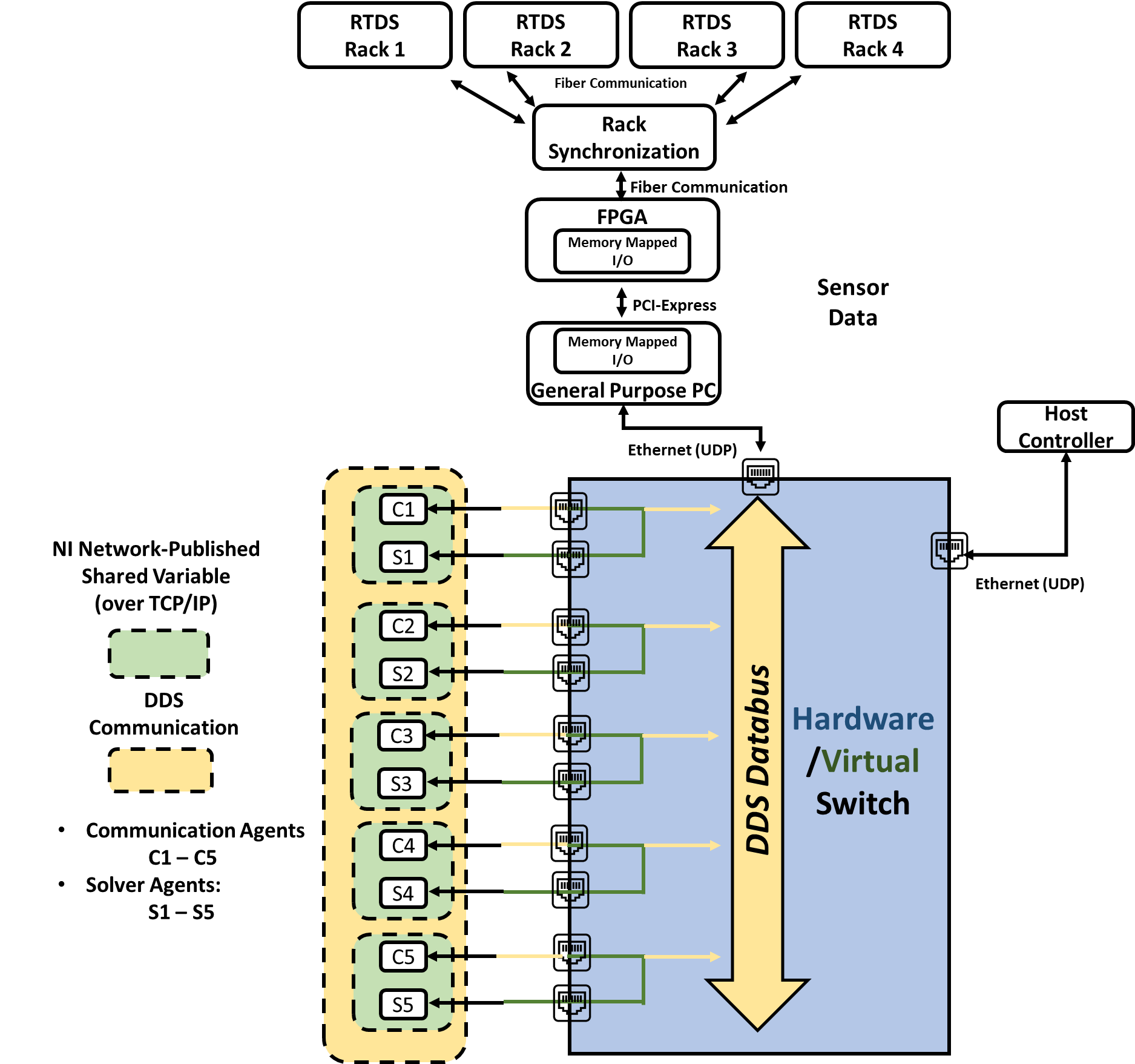}}
\caption{Overview of the experimental setup used to evaluate the performance of the distributed energy management control system for MVDC ship power systems.}
\label{fig:overall_diagram}
\vspace{-5mm}
\end{figure}

In this section, we provide a description of the methodology used to develop the CEF communication network module and the experimental setup used to capture the network characteristics critical to perform effective control evaluations under the CEF. The proposed communication network module is designed to provide an effective way of testing multiple non-ideal communication network configurations under the CEF. Fig. \ref{fig:cef_cyber} shows the version of the CEF with the addition of the communication network module proposed. As seen in this figure, the proposed module is made up of resources and experimental setups tailored to provide support for communication network evaluations of control systems. In the following subsections, the CEF communication network module will be used to replicate the behavior of an `ideal' hardware-based communication network deployed to test the performance of a distributed power and energy management system for MVDC ship power systems \cite{esm_mvdc}. Fig. \ref{fig:overall_diagram} shows a detailed diagram of the experimental setup used to deploy the distributed power and energy management system. 

\subsection{Hardware-based Communication Network of Distributed Energy Management System for MVDC Ship Power System
}
\begin{figure*}[ht]
\centerline{\includegraphics[width=12.7cm, height=8.6cm]{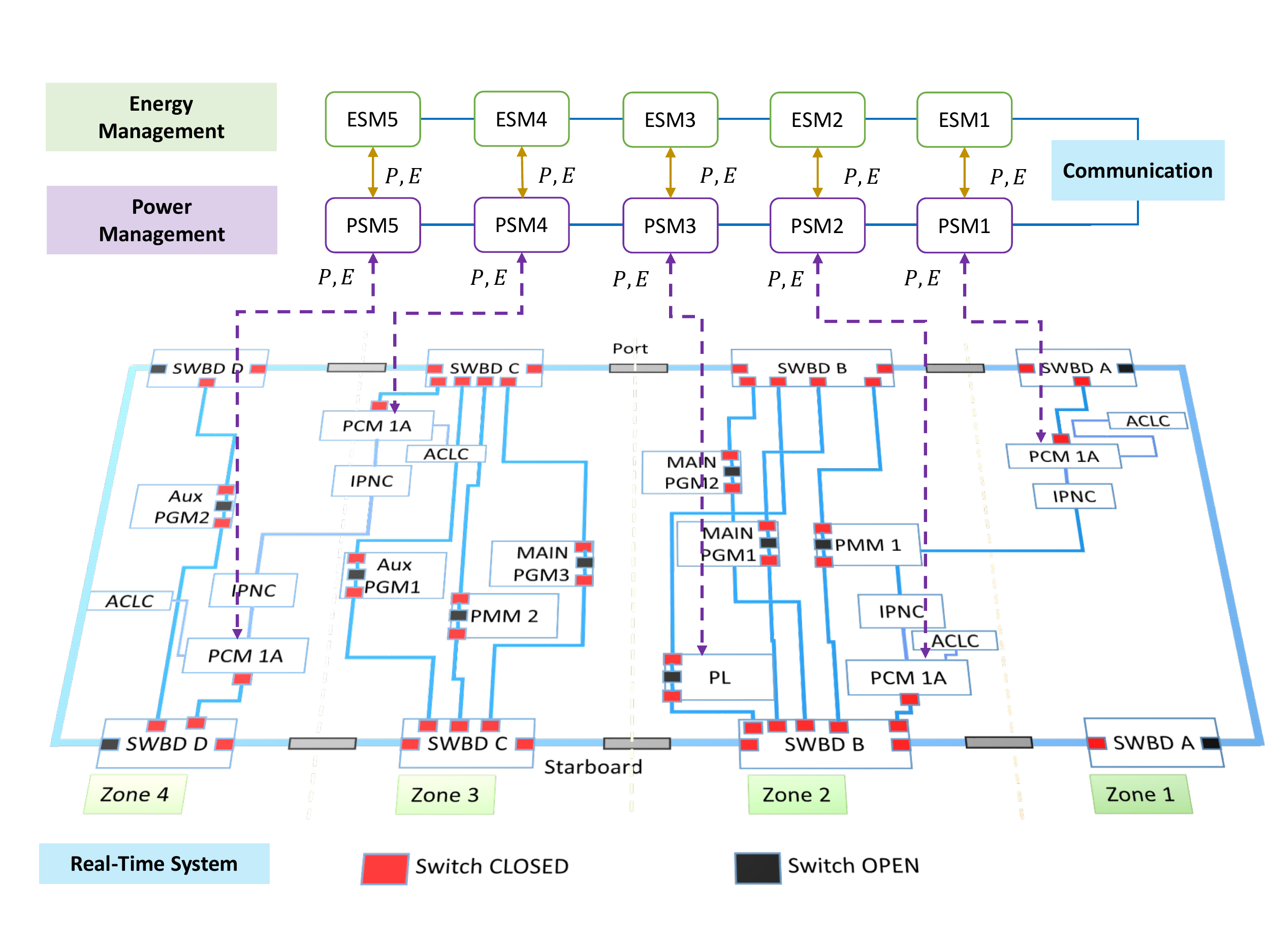}}
\caption{4-Zone MVDC and Energy Storage Control Functions. Note: $P$, $E$ are the power and energy signals.}
\label{fig:Control Architecture}
\end{figure*}

The distributed energy storage management system taken into account for evaluating the effect of the investigated network environment is shown in Fig. \ref{fig:Control Architecture}. In the figure, the test system is a 4-zone MVDC system \cite{4-Zone_MVDC_Model}. The power system is a 12kVDC-100MW MVDC distribution system implemented in 4 simulation racks of the RTDS. In the model, there are five distributed energy storage modules (ESM) residing in the 1kV distribution system of 4 PCM-1A and Pulse Load (PL). Each ESM is capable of discharging 10MW and charging 5MW power. The maximum capacity of each ESM is 1GJ. In our test cases, these ESM will begin with their stage of charge (SOC) of 50\% and will be charged to a predefined SOC of 80\% before going to serve the ship missions. The predefined 80\% will be kept as the set-point for the ESM module during ship operations. These ESMs are controlled under a distributed power and energy management system scheme. We have presented our control system for the energy storage modules in \cite{Vu2018_largescale}. There are also five power management controllers for ESM (PSM1-PSM5), which manage the power-sharing among the energy storage module during pulse power operations with a spinning reserve operation mode. On the other hand, the other five energy management controllers for ESM (ESM1-ESM5) manage the charging operation of the energy storage to a predefined SOC upon the completion of every pulse power operation. In this control architecture, the PSM controllers communicate with each other in a ring topology with the communication bandwidth of 1$ms$ while each ESM controller communicates with each other in a star topology with a communication bandwidth of 5$ms$. In this paper, we focus on evaluating the communication impacts (including delays and package drops) on the ESM control performance.
    
In relation to the communication network, every ESM controller is made up of a communication agent (C1-C5) and a solver agent (S1-S5). The communication agents (C1-C5) communicate with each other through a DDS publish-subscribe model while each solver agent (S1-S5) communicates with its corresponding communication agent, i.e., C1 $\leftrightarrow$ S1, C2 $\leftrightarrow$ S2, ... , C5 $\leftrightarrow$ S5, using NI network-published shared variable over TCP/IP. Fig. \ref{fig:overall_diagram} shows the DDS communication flow represented by the yellow lines and the NI network-published shared variable communication flow represented by the green lines. The current implementation of the distributed energy storage management system is performed using a hardware-based switch that facilitates the communication between all the system's nodes.

\vspace{-1mm}
\subsection{Virtual-based Communication Network of Distributed Energy Management System for MVDC Ship Power System}

\begin{figure}[ht]
\centerline{\includegraphics[width=0.35\textwidth]{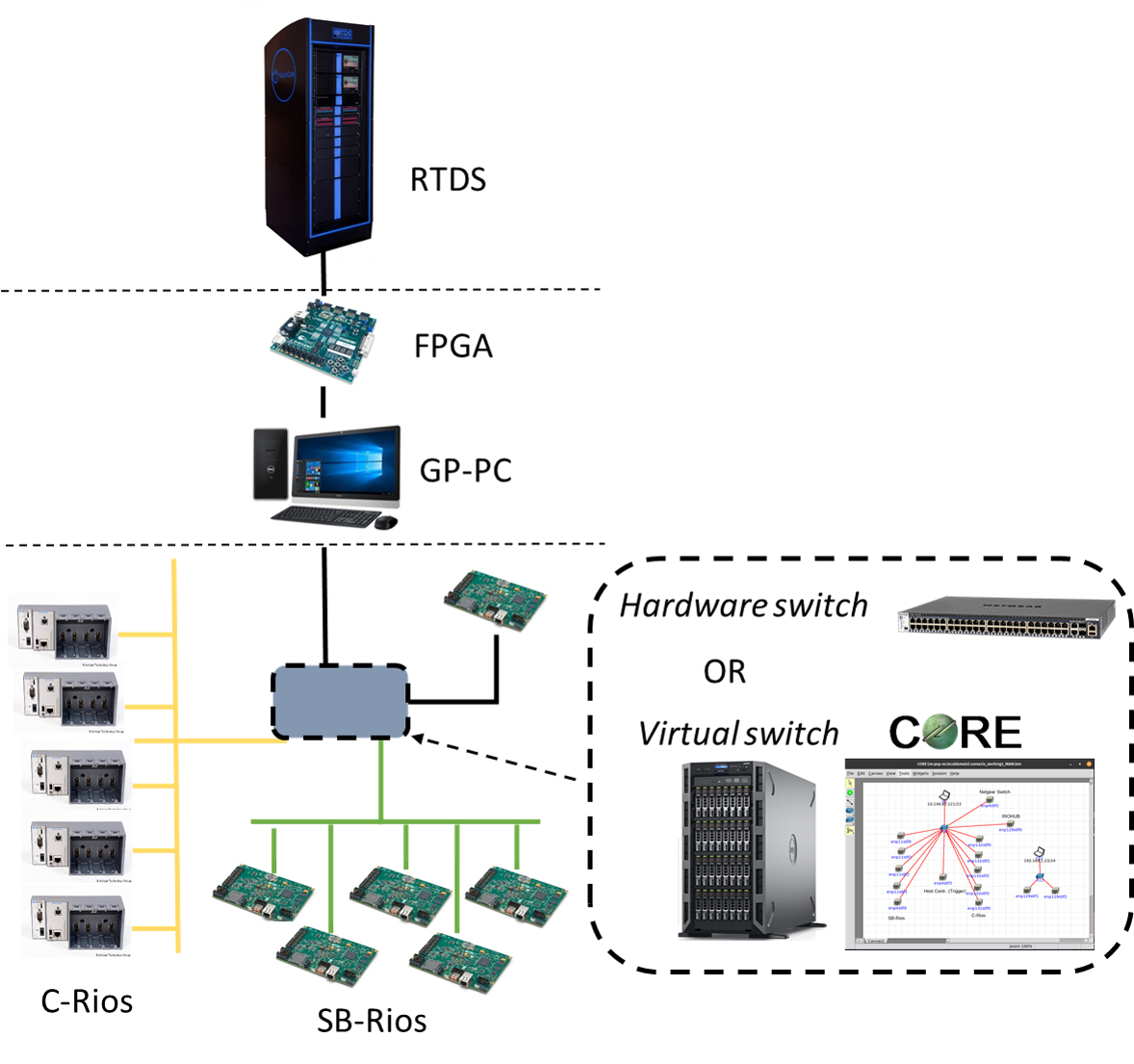}}
\caption{Overview of hardware components present in the experimental setup used to evaluate the performance of the distributed energy management control system for MVDC ship power systems.}
\label{fig:hardware_diagram}
\end{figure}

\begin{figure}[ht]
\centerline{\includegraphics[width=0.24\textwidth]{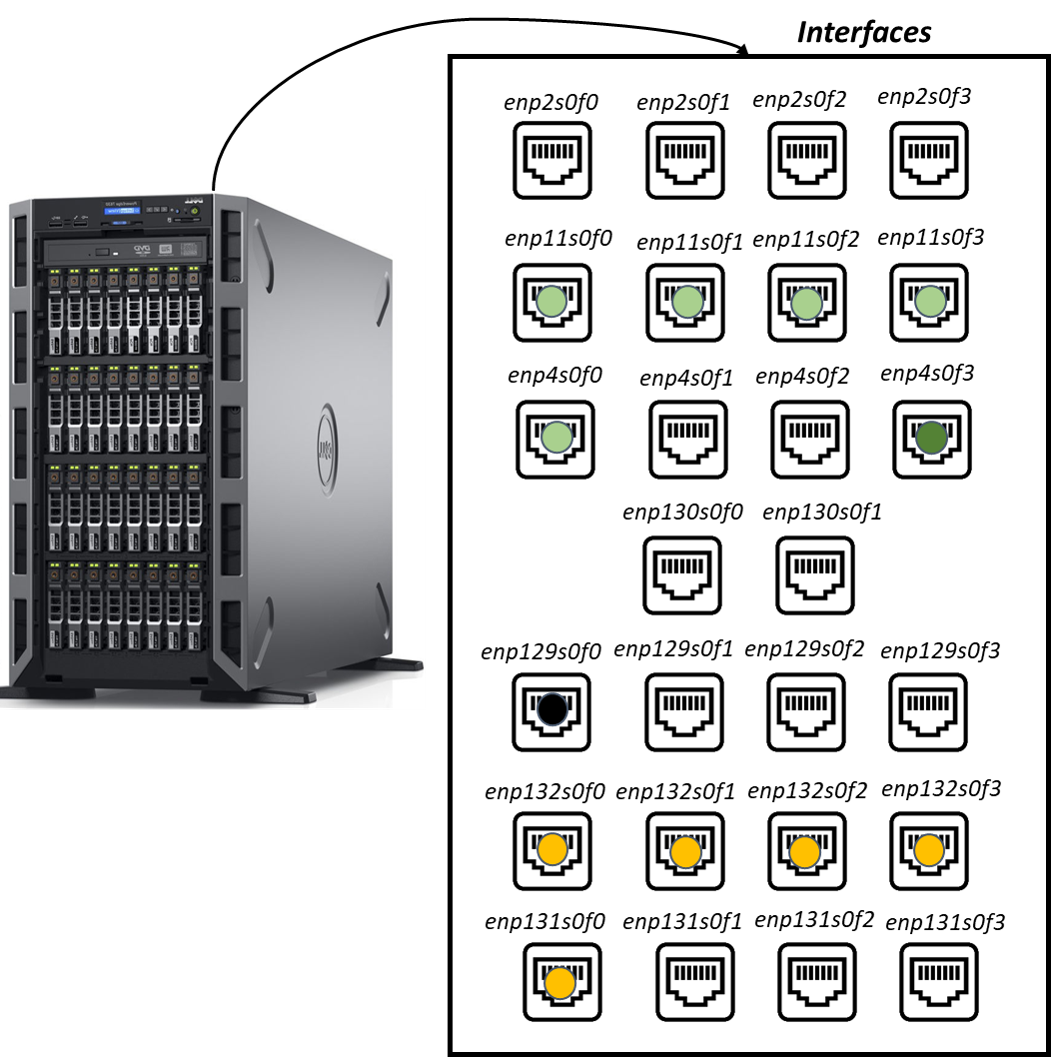}}
\caption{Diagram with physical connections to the high-performance server running CORE emulation software.}
\label{fig:connections}
\end{figure}

In order to replicate the behavior of the `ideal' hardware-based communication network discussed above, a virtual model of the communication network was modeled using the Common Open Research Emulator (CORE) \cite{corenetwork} software and deployed in a high-performance server with support for multiple network interface cards (NICs). Fig. \ref{fig:hardware_diagram} shows an overall diagram with all the hardware components mapped one-to-one to the blocks shown in Fig. \ref{fig:overall_diagram}. As seen, the virtual switch and network links are emulated inside the high-performance server and all the C-RIOs and SB-RIOs controllers are connected to individual NICs. Fig. \ref{fig:connections} shows a detailed diagram of how these connections were implemented. The yellow dots represent the connections to the five communication agents (c-RIOs) and the light green dots represent the connections to the five solver agents (SB-RIOs). The dark green dot represents the connection to the Host controller and the black dot represents the connection to the general-purpose PC (GP-PC) in charge of relaying the sensor data coming to and from the physical-system simulated inside the RTDS. Each one of these physical connections is bridged to an RJ45 virtual node inside CORE. More details about this process and the CORE architecture are presented below.

\subsection{Common Open Research Emulator (CORE) Architecture}
The Common Open Research Emulator (CORE) is an open-source emulation software that provides high-fidelity emulation for communication networks. CORE is capable of emulating different types of routers, Ethernet switches, Ethernet hubs, wireless local area networks (WLANs), and virtual PCs. CORE emulates the application, presentation, session, transport, and network OSI layers through Linux network namespacing virtualization and uses simplified simulated links to simulate the behavior of the data link and physical link layers \cite{ahrenholz2010comparison}. The simplified link models used for the data link and physical layers are realized using netem \cite{netem} with bridging in Linux environments. The use of netem allows the emulation of statistical network effects such as percentage packet loss, bandwidth, and delays. On the other hand, the virtualization of the network and above OSI layers and operational hosts is realized using Linux name-spacing (i.e., \textit{netns}) for creating the virtual nodes for the specific devices and then are linked together using Linux bridging and virtual interfaces. CORE uses a paravirtualization technique, where only part of the operating system (OS) is virtualized since only processes and network stacks are isolated, while hardware such as disks, timers, and other devices are shared between the defined nodes. This approach of only virtualizing what is necessary for the network emulation makes CORE a very lightweight high-fidelity emulation software that can be scaled to over a hundred virtual network devices running in a single emulation machine. Detailed evaluations of the performance of CORE on different platforms are presented in \cite{realtime_core, integration_core}.

Due to CORE's capability of running in real-time, real network devices and equipment such as routers, switches, hubs, and controllers can be connected so that they can interact in real-time with the virtual communication networks modeled inside CORE. This capability is the main driver for the use of CORE as the software platform for the CEF communication network module.

\vspace{-1mm}
\section{Case Studies and Results}
This section presents the results and performance evaluation of the virtual communication network model compared to the hardware-based `ideal' communication network model. To evaluate the performance of the experimental setup, two benchmark scenarios are implemented: a) response times of the hardware-based switch vs. virtual-based switch, and b) ESM controllers' response to statistical network effects. The hardware-based communication network is realized using an off-the-shelf commercial Netgear M4300-52G switch. The virtual communication network is running in a high-performance Dell server with Intel Xeon CPU E5-2637 v4 with a clock rate of 3.50 GHz, 64 GB of RAM, and 26 i350 Gigabit network connections. 

\subsection{Benchmark 1: Response Times Through Hardware-based and Virtual-based Switch}

The following benchmarks intend to provide confidence that the computer network emulation can faithfully represent 10's (up to approximately 100) of network-connected control nodes. The following benchmark has yet to be executed at such a scale, but the initial results do show promise for the network emulation implementation.  

A typical control can be modeled as a periodic task and therefore this benchmark intends to replicate such behavior as seen from the perspective of the computer network. More specifically, an \emph{echo} benchmark test was performed to compare the timing characteristics of the virtual communication network compared to the hardware-based communication network. The test performed consisted of measuring the response times for UDP packet exchanges between a periodic sender and an echo server. In this \emph{echo} benchmark, every elapsed period
of time the sender sends a UDP packet to the echo server. Upon receiving the sender's packet the echo server immediately returns the received packet back to the sender. The response time is calculated as the length of time between the time instant the sender sends the packet to the time instant the sender receives the echoed packet.

\begin{figure}[h]
    \centering
    \includegraphics[width=0.95\columnwidth]{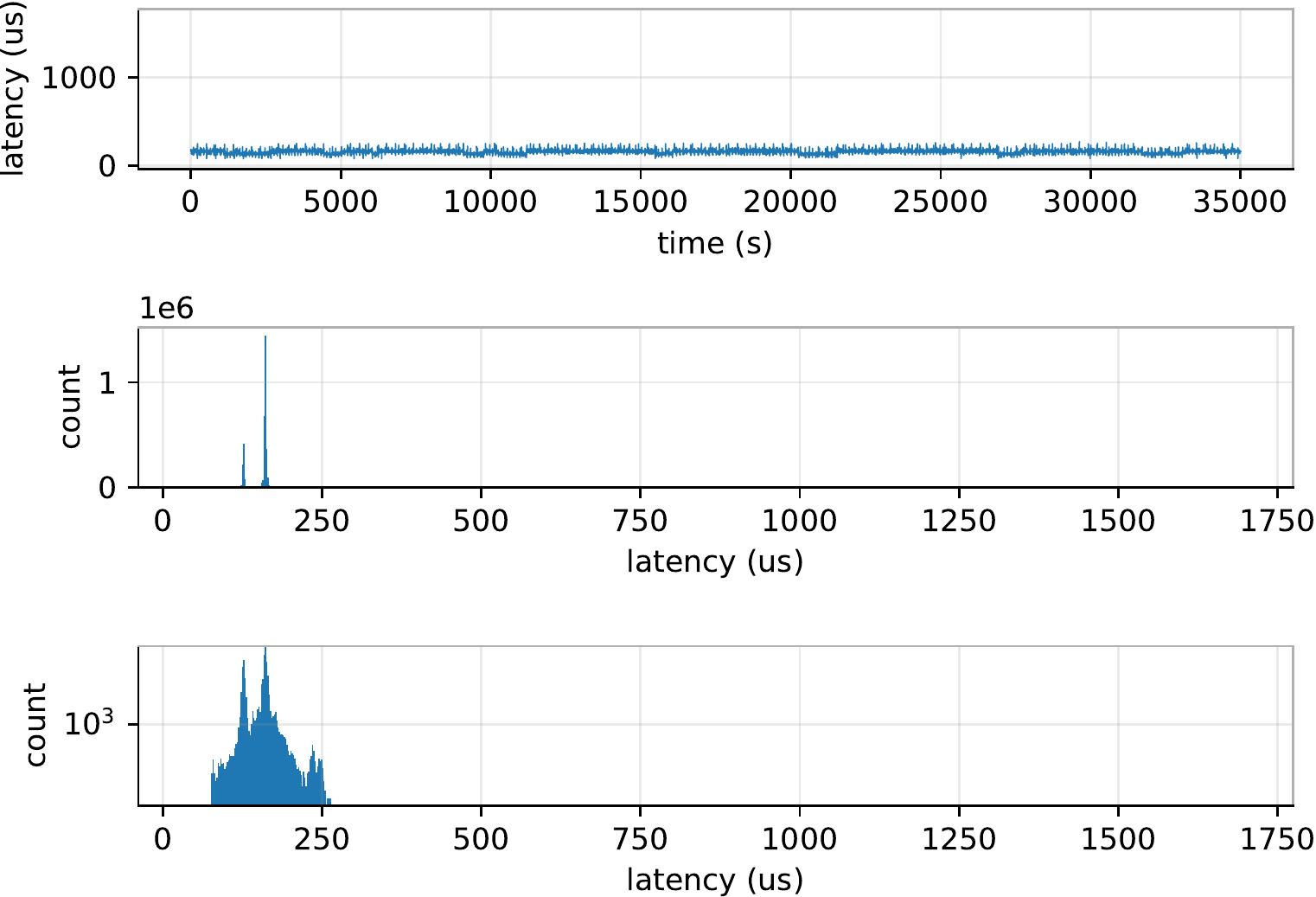}
    \caption{Response times for \emph{hardware-based} switch (\emph{Netgear M4300-52G}).}
    \label{fig:real-switch-response-times}
\end{figure}

\begin{figure}[h]
    \centering
    \includegraphics[width=0.95\columnwidth]{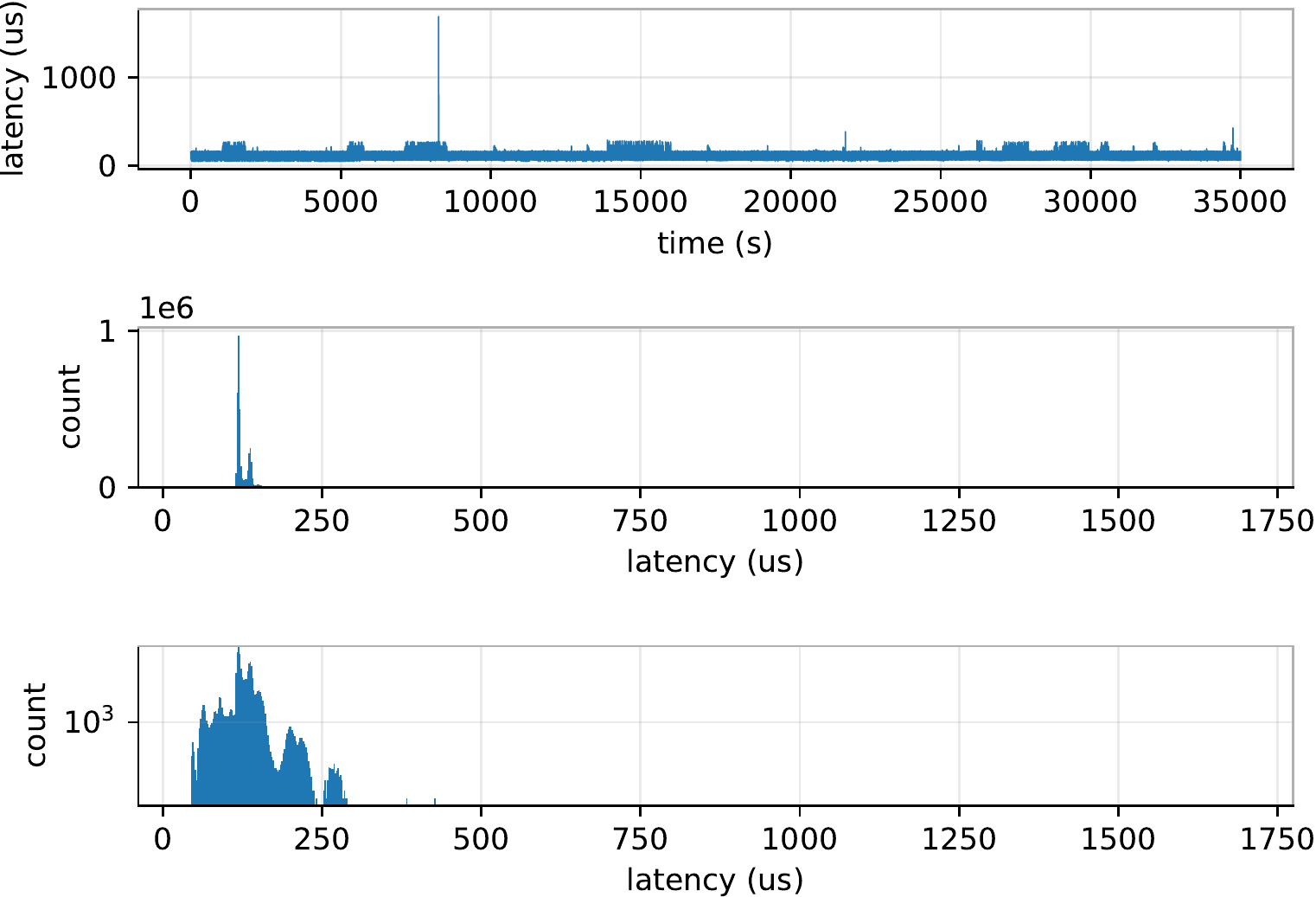}
    \caption{Response times for \emph{virtual-based} switch.}
        \label{fig:virtual-switch-response-times}
\end{figure}

The response times for an echo benchmark with a period of $\sim$10 $ms$ are plotted for the hardware switch in Fig. \ref{fig:real-switch-response-times} and for the virtual-based switch (modeled in CORE) in Fig. \ref{fig:virtual-switch-response-times}. The topmost plot in the response time figures is the individual response times for each packet sent by the sender for the duration of experiment. The bottom two plots illustrate the distribution of response times. These bottom two plots are identical except that the bottom-most plot has the response time measurement distributions plotted on a log scale. As seen in the figures, the differences between the hardware-based implementation and the virtual-based implementation are negligible for applications with communication requirements in the range of 100 to 250 $\mu s$ and above. 

The majority of response times are comparable between the hardware- and virtual-based switch. Surprisingly, the response times of the virtual switch generally show lower latencies than that of the hardware switch; the virtual switch results in a mean of 124 $\mu s$ whereas the hardware switch results in a mean of 153 $\mu s$. The virtual switch does have a few unexpectedly long response times but these long response times are relatively few. These outliers are being analyzed and are expected to be alleviated by configuring the kernel running the virtual switch for more deterministic response times rather than the default configuration of low average-case performance.

It is important to note that the measured response times include time waiting for packet processing in the OS and are not limited to time for propagating data through the computer network. That being said, a typical system-level control will likely execute on a computer with an OS and therefore the response times can be considered representative of the communication time as seen from the perspective of the control process. In addition, these tests also demonstrate other advantages related to the capabilities of modeling more complex communication network topologies that would have a similar behavior to a real communication network.

\subsection{Benchmark 2: ESM Controllers' Response}

Controllers' response benchmark tests are conducted to evaluate the response of the ESM controllers depending on which communication network, hardware-based or virtual-based, the controllers are connected to. Additional tests were conducted to investigate and evaluate the response of the controllers to added \textit{artificial} statistical network effects such as increased delays and percentage packet drops in the communication network. The specific case studies conducted to evaluate the controllers' responses are presented below.

\begin{enumerate}
    \item \textit{ESM controllers' response using hardware-based communication network.}
    \item \textit{ESM controllers' response using virtual-based communication network.}
    \item \textit{ESM controllers' response using virtual-based communication network with 10ms delays.}
    \item \textit{ESM controllers' response using virtual-based communication network with 100ms delays.}
    \item \textit{ESM controllers' response using virtual-based communication network with 10\% packet drops.}
\end{enumerate}

It should be noted that the response of the distributed energy management system differs slightly at every run even when using the exact same testing data. So, in order to perform a fair comparison between case studies, statistical metrics were used to evaluate the percentage differences between the different test runs of each case study and between the case studies themselves. The SOC of each ESM is the variable used to characterize the controller's response for each case study and each test run. The two main metrics used for the metrics analysis are the mean absolute percentage error (MAPE) and the average percentage difference (PD) for each ESM SOC at each test run. The equations of the metrics are shown below.

\begin{equation}
    MAPE (\%) = \frac{1}{n}\sum_{i=1}^{n} \Bigl|\frac{x_1^i - x_2^i}{x_1^i}\Bigr| \cdot 100
\end{equation}

\begin{equation}
    PD_i (\%) = \frac{x_1^i - x_2^i}{\frac{1}{2}(x_1^i + x_2^i)} \cdot 100
\end{equation}

\noindent Where $x_1$ and $x_2$ represent the two time-series signals being compared and $n$ is the total sample size of the signals. Table \ref{tab:stats_of_runs} shows the $MAPE(\%)$ and average $PD(\%)$ for each ESM SOC signal compared between test runs in each case study. As observed in these results, the percentage differences (variations) of the controllers' responses between test runs are not significant enough to limit our capability of comparing different test case study scenarios. Table \ref{tab:stats_of_cases} shows the $MAPE(\%)$ and average $PD(\%)$ for each ESM SOC signal compared between runs \# 1 for each case study. Here, we can observe how the network effects significantly affect the performance of the distributed energy management system.

\begin{table}[]
\centering
\setlength{\tabcolsep}{3.5pt}
\caption{Comparison of test runs \#1 and \#2 for each case study.}
\label{tab:stats_of_runs}
\begin{tabular}{||c|c|c|c|c||}
\hline
\textbf{\begin{tabular}[c]{@{}c@{}}Case Study\\ Scenario\end{tabular}} & \textbf{\begin{tabular}[c]{@{}c@{}}Test\\ Runs\end{tabular}} & \textit{\textbf{SOC}} & \textit{\textbf{MAPE (\%)}} & \textit{\textbf{Avg. PD (\%)}} \\ \hline \hline
\multirow{5}{*}{\begin{tabular}[c]{@{}c@{}}Hardware-based \\ Switch\end{tabular}} & \multirow{5}{*}{\begin{tabular}[c]{@{}c@{}}Run \\ 1 vs. 2\end{tabular}} & ESM 1 & 0.045 & 0.045 \\ \cline{3-5} 
 &  & ESM 2 & 0.065 & 0.065 \\ \cline{3-5} 
 &  & ESM 3 & 0.056 & 0.056 \\ \cline{3-5} 
 &  & ESM 4 & 0.042 & 0.042 \\ \cline{3-5} 
 &  & ESM 5 & 0.155 & 0.111 \\ \hline \hline
\multirow{5}{*}{\begin{tabular}[c]{@{}c@{}}Virtual-based\\  Switch\end{tabular}} & \multirow{5}{*}{\begin{tabular}[c]{@{}c@{}}Run \\ 1 vs. 2\end{tabular}} & ESM 1 & 0.034 & 0.034 \\ \cline{3-5} 
 &  & ESM 2 & 0.030 & 0.034 \\ \cline{3-5} 
 &  & ESM 3 & 0.032 & 0.041 \\ \cline{3-5} 
 &  & ESM 4 & 0.036 & 0.041 \\ \cline{3-5} 
 &  & ESM 5 & 0.136 & 0.153 \\ \hline \hline
\multirow{5}{*}{\begin{tabular}[c]{@{}c@{}}Virtual-based\\  Switch\\ 10 $ms$ delay\end{tabular}} & \multirow{5}{*}{\begin{tabular}[c]{@{}c@{}}Run \\ 1 vs. 2\end{tabular}} & ESM 1 & 0.034 & 0.011 \\ \cline{3-5} 
 &  & ESM 2 & 0.045 & 0.025 \\ \cline{3-5} 
 &  & ESM 3 & 0.041 & 0.045 \\ \cline{3-5} 
 &  & ESM 4 & 0.041 & 0.041 \\ \cline{3-5} 
 &  & ESM 5 & 0.153 & 0.039 \\ \hline \hline
\multirow{5}{*}{\begin{tabular}[c]{@{}c@{}}Virtual-based\\  Switch\\ 100 $ms$ delay\end{tabular}} & \multirow{5}{*}{\begin{tabular}[c]{@{}c@{}}Run \\ 1 vs. 2\end{tabular}} & ESM 1 & 0.170 & 0.170 \\ \cline{3-5} 
 &  & ESM 2 & 0.242 & 0.240 \\ \cline{3-5} 
 &  & ESM 3 & 0.169 & 0.169 \\ \cline{3-5} 
 &  & ESM 4 & 0.122 & 0.122 \\ \cline{3-5} 
 &  & ESM 5 & 0.427 & 0.425 \\ \hline \hline
\multirow{5}{*}{\begin{tabular}[c]{@{}c@{}}Virtual-based\\  Switch\\ 10\%\\ packet drop\end{tabular}} & \multirow{5}{*}{\begin{tabular}[c]{@{}c@{}}Run \\ 1 vs. 2\end{tabular}} & ESM 1 & 0.452 & 0.458 \\ \cline{3-5} 
 &  & ESM 2 & 0.421 & 0.420 \\ \cline{3-5} 
 &  & ESM 3 & 0.518 & 0.518 \\ \cline{3-5} 
 &  & ESM 4 & 0.371 & 0.372 \\ \cline{3-5} 
 &  & ESM 5 & 0.705 & 0.695 \\ \hline \hline
\end{tabular}
\end{table}

\begin{figure*}[t]\centering
\subfigure[] { \label{fig:hsw}     
\includegraphics[width=4.32cm]{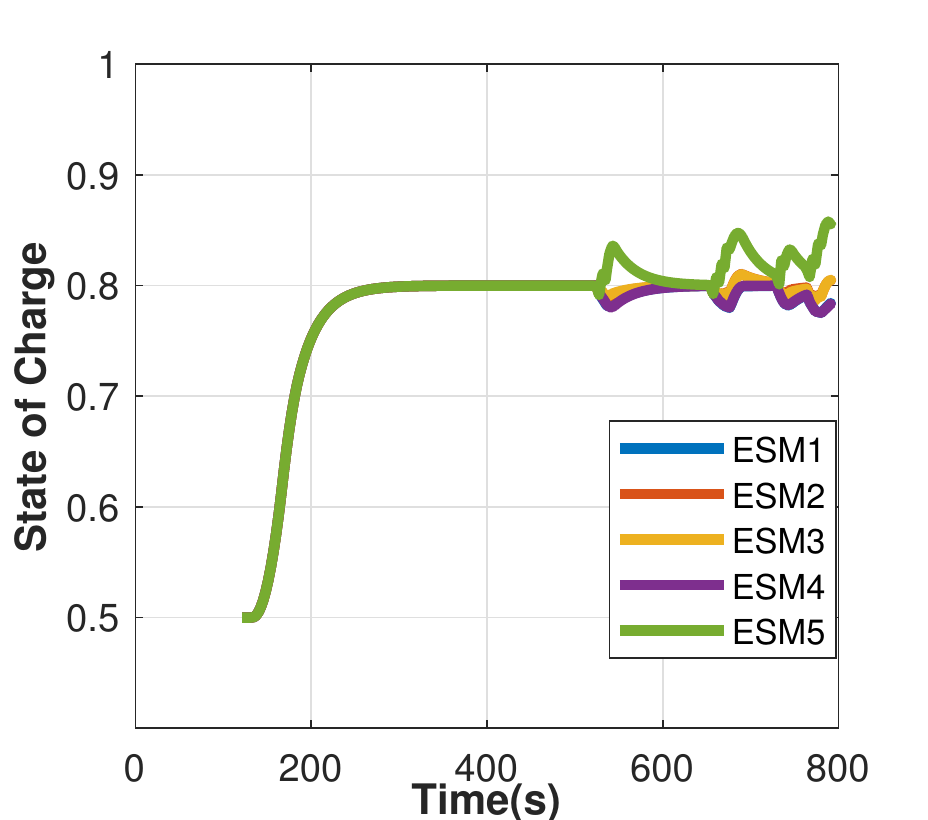}}
\subfigure[] { \label{fig:vsw}     
\includegraphics[width=4.32cm]{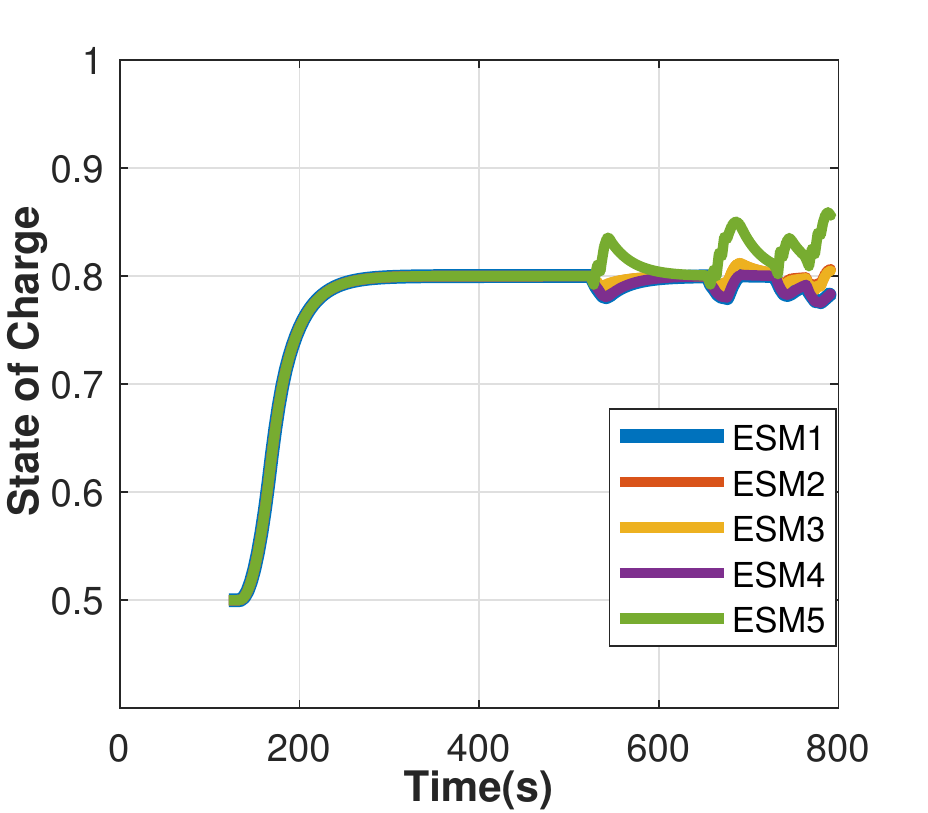}}
\subfigure[] { \label{fig:vsw100}     
\includegraphics[width=4.32cm]{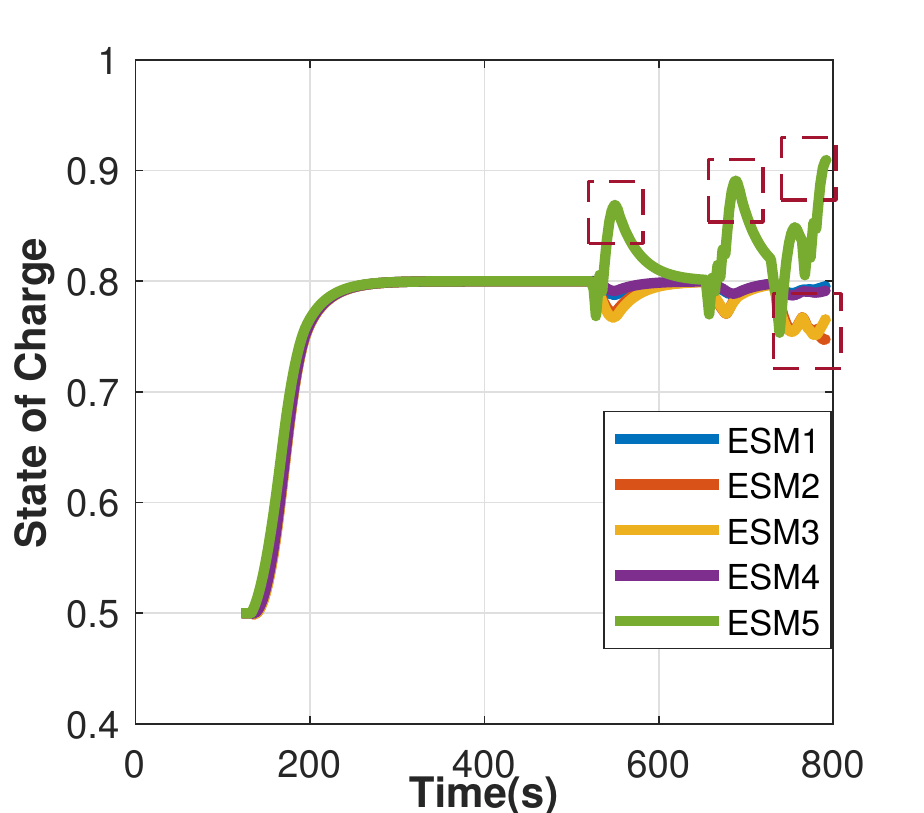}}
\subfigure[] { \label{fig:vsw10perc}     
\includegraphics[width=4.32cm]{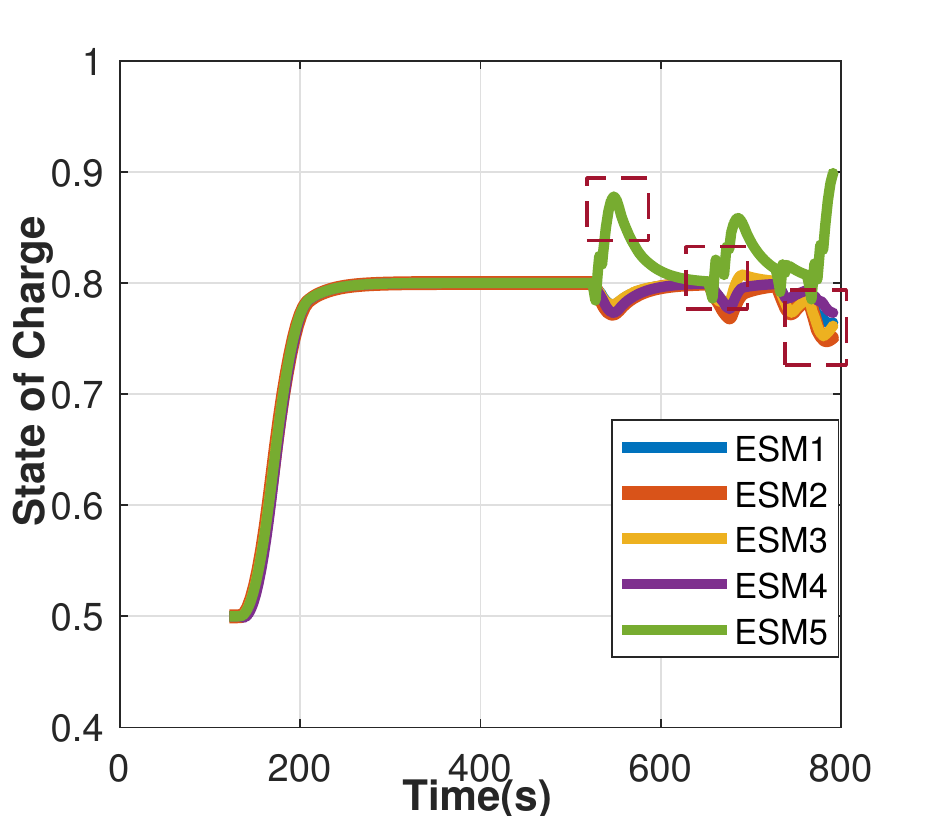}}
\caption{SOC graphs of ESM1-ESM5 during run \#1 of each case study scenario: \subref{fig:hsw} hardware-based switch communication network, \subref{fig:vsw} virtual-based switch communication network, \subref{fig:vsw100} virtual-based switch communication network with 100 $ms$ delay, and \subref{fig:vsw10perc} virtual-based switch communication network with 10\% packet drop.}
\label{fig:results_graphs}
\vspace{-4mm}
\end{figure*}

Figs. \ref{fig:results_graphs} depict the SOC of the five ESMs of the shipboard power system in a single run of the distributed energy management system for each case study implemented. Fig. \ref{fig:hsw} shows the SOC of all the ESMs when the hardware-based switch communication network is used. This case study serves as a baseline representation of the typical characteristics observed when testing controls in the CEF. Fig. \ref{fig:vsw} depicts the SOC of the ESMs for case study two, i.e., when the hardware-based switch is substituted with the virtual-based switch emulated within CORE. The compared variation between hardware- and virtual-based switches are difficult to visualize but are consistent with the numerical data reported in Table \ref{tab:stats_of_runs}. Fig. \ref{fig:vsw100} demonstrates the case study involving the virtual-based switch from a single run, where a 100$ms$ communication delay is imposed on the communication network. The impact of the imposed communication delay shows an increased SOC magnitude on all ESM controllers when compared to either the hardware- or virtual-based switch case studies. The red boxes in the graph show the sections were the majority of the differences can be observed. The delay imposed in the network causes a delay in the consensus setpoints that the distributed controllers shared between them; thus making the system behave erroneously.

A similar result was obtained when the communication network was modified to account for a 10\% packet drop in the network. Fig. \ref{fig:vsw10perc} presents observations obtained from another single-run use-case utilizing the virtual-based switch, where a 10\% packet loss is imposed on ESM controllers. Here, it can also be observed that the controllers take more time to reach consensus and have a higher number of oscillations when reaching the desired set points. 

\begin{table}[]
\centering
\setlength{\tabcolsep}{3.5pt}
\caption{Comparison between case studies' runs \#1.}
\label{tab:stats_of_cases}
\begin{tabular}{||c|c|c|c||}
\hline
\textbf{\begin{tabular}[c]{@{}c@{}}Case Study\\ Scenario\end{tabular}} & \textit{\textbf{SOC}} & \textit{\textbf{MAPE (\%)}} & \textit{\textbf{Avg. PD (\%)}} \\ \hline \hline
\multirow{5}{*}{\begin{tabular}[c]{@{}c@{}}Hardware-based \\ Switch Run \#1\\ vs.\\ Virtual-based\\ Switch Run \#1\end{tabular}} & ESM 1 & 0.021 & 0.034 \\ \cline{2-4} 
 & ESM 2 & 0.028 & 0.091 \\ \cline{2-4} 
 & ESM 3 & 0.025 & 0.010 \\ \cline{2-4} 
 & ESM 4 & 0.018 & 0.004 \\ \cline{2-4} 
 & ESM 5 & 0.046 & 0.049 \\ \hline \hline
\multirow{5}{*}{\begin{tabular}[c]{@{}c@{}}Virtual-based \\ Switch Run \#1\\ vs.\\ Virtual-based Switch \\ - 10 $ms$ delay Run \#1\end{tabular}} & ESM 1 & 0.225 & 0.564 \\ \cline{2-4} 
 & ESM 2 & 0.509 & \textbf{1.666} \\ \cline{2-4} 
 & ESM 3 & 0.504 & \textbf{1.607} \\ \cline{2-4} 
 & ESM 4 & 0.219 & 0.590 \\ \cline{2-4} 
 & ESM 5 & 0.965 & \textbf{4.779} \\ \hline \hline
\multirow{5}{*}{\begin{tabular}[c]{@{}c@{}}Virtual-based \\ Switch Run \#1\\ vs.\\ Virtual-based Switch \\ - 100 $ms$ delay Run \#1\end{tabular}} & ESM 1 & 0.394 & 0.727 \\ \cline{2-4} 
 & ESM 2 & 0.921 & \textbf{3.731} \\ \cline{2-4} 
 & ESM 3 & 0.951 & \textbf{2.392} \\ \cline{2-4} 
 & ESM 4 & 0.405 & 0.491 \\ \cline{2-4} 
 & ESM 5 & 1.150 & \textbf{3.114} \\ \hline \hline
\multirow{5}{*}{\begin{tabular}[c]{@{}c@{}}Virtual-based \\ Switch Run \#1\\ vs.\\ Virtual-based Switch \\ - 10\% packet drop Run \#1\end{tabular}} & ESM 1 & 0.349 & 1.249 \\ \cline{2-4} 
 & ESM 2 & 0.838 & \textbf{3.418} \\ \cline{2-4} 
 & ESM 3 & 0.610 & \textbf{2.692} \\ \cline{2-4} 
 & ESM 4 & 0.491 & 0.692 \\ \cline{2-4} 
 & ESM 5 & 0.774 & \textbf{2.613} \\ \hline \hline
\end{tabular}
\end{table}

\section{Discussion and Conclusions}
Initial experimental observations pertaining to the implementation of the real-time communication network module within the CEF are very promising. Modification of the existing experimental setup to incorporate the virtual-based switch was successfully accomplished, requiring the substitution of a single component (hardware-based switch). An initial expectation of this substitution was an increase in signal communication times, however, experimental observations pertaining to the virtual-based switch depict similar signal communication times when compared to the hardware-based switch. Future testing will investigate the consistency of this observation with respect to the addition of larger numbers of hardware controllers. The explored use cases permitted the instantiation of non-ideal communication characteristics, a primary motivator for this endeavor. Both the addition of signal delay and packet loss demonstrated an expected deterioration in the controllers' performance. These observations are important because they help validate the implementation of this module as a useful tool for exploring control performance under a variety of communication conditions. While results obtained using CORE are promising, there are some software-related limitations pertaining to the available feature set. Future work will explore other software packages to increase the available configuration options.

\vspace{-1mm}
\bibliographystyle{IEEEtran}
\bibliography{bibliography/main}

\end{document}